\documentstyle[11pt,paspconf,epsf,psfig,twoside]{article}

\newcommand\lax{{\lower0.75ex\hbox{$<$}\atop\raise0.5ex\hbox{$\sim$}}}
\newcommand\gax{{\lower0.75ex\hbox{$>$}\atop\raise0.5ex\hbox{$\sim$}}}
\newdimen\digitwidth
\setbox0=\hbox{\rm0}
\digitwidth=\wd0
\catcode`!=\active
\def!{\kern\digitwidth}
\begin{document}

\title{EUVE Spectroscopy of Polars}

%FOR MY USE ONLY%%%%%%%%%%%%%%%%%%%%%%%%%%%%%%%%%%%%%%%%%%%%%%%%%%%%%%%%%%%%%%%%%%%%%%%%%%%
\vbox to -10pt{\vskip -3.0cm
\hbox to \hsize{%
	{\it 1998,  Proceedings of the Annapolis Workshop on Magnetic Cata-}%
	{\hbox to 1.3cm{\hfil }}}
\hbox to \hsize{%
	{\it clysmic Variables, ed.\ C.~Hellier \& K.~Mukai (San Francisco: ASP)}%
	{\hbox to 1.3cm{\hfil }}}
\vss}
\vskip -10pt
%%%%%%%%%%%%%%%%%%%%%%%%%%%%%%%%%%%%%%%%%%%%%%%%%%%%%%%%%%%%%%%%%%%%%%%%%%%%%%%%%%%%%%%%%%%

\author{Christopher W.\ Mauche}
\affil{Lawrence Livermore National Laboratory, CA, U.S.A.\ and\\
       Department of Physics, Keele University, Staffordshire, U.K.}

\begin{abstract}
An admittedly pedantic but hopefully useful and informative analysis is presented of
the {\it EUVE\/} 70--180~\AA \ spectra of nine polars. These spectra are fit with
three different models---a blackbody, a pure-H stellar atmosphere, and a solar
abundance stellar atmosphere---to reveal the presence of spectral features such as
absorption lines and edges, and to investigate the sensitivity of the derived ($kT$,
$N_{\rm H}$, solid angle) and inferred (fractional emitting area, bolometric luminosity)
parameters to the model assumptions. Among the models tested, the blackbody model best
describes the observed spectra, although the untested irradiated solar abundance stellar
atmosphere model is likely a better overall description of the EUV/soft X-ray spectra
of polars.
\end{abstract}


\section{Introduction}

When all is said and done, the accreting material that causes all the fireworks in a
magnetic CV finds itself channeled onto a small spot on the white dwarf surface in the
vicinity of the magnetic pole(s). The extremes of this accretion region are masked from us
by the units we typically employ to describe it: translated into more familiar units, a
shock temperature of 10 keV equals 100 million degrees; an accretion rate of $10^{-10}\>
\rm M_\odot~yr^{-1}$ equals 7 billion tons per second; an accretion luminosity of $10^{33}
\> \rm ergs~s^{-1}$ is the equivalent energy release of $2\times 10^{10}$ megaton bombs
per second. All this energy is released above and into an area of $\sim 10^{-3}$ times the
surface area of the white dwarf---an area of $\sim 200\, 000~\rm miles^2$, which is about
the size of California.

The energy input into the accretion region is supplied by radiative heating from above by
the $\sim 10$~keV thermal plasma below the accretion shock and by mechanical heating by
blobs of material which punch through the shock and penetrate into the stellar surface
before thermalizing their kinetic energy. The equilibrium photospheric temperature of the
region is then determined by the balance of radiative and mechanical heating and radiative
cooling, with the latter dependent upon such factors as the surface area of the accretion
region and the sources of opacity (i.e., metallicity) and the ionization state of the
photosphere. For a luminosity of $10^{33}~\rm ergs~s^{-1}$ and a fractional emitting area
of $\sim 10^{-3}$, the blackbody temperature of the region is $\sim 20$~eV.

Unfortunately, it is observationally challenging to accurately determine the spectral
parameters of a $\sim 20$~eV blackbody: its peak (in $dE/d\lambda $) lies at $\sim 100$
\AA\ or $\sim 0.1$~keV where the energy resolution of ionization-type detectors is poor
and photoelectric absorption is severe. Worse, dispersive instruments do not give
consistent results for AM~Her, by far the brightest polar: the best-fit parameters
of a blackbody fit to the {\it Einstein\/} OGS spectrum of AM~Her are $kT=46$~eV and
$N_{\rm H}=3.2\times 10^{19}~\rm cm^{-2}$ (Heise et al.\ 1984), the parameters for the
{\it EXOSAT\/} TGS spectrum are $kT=28$~eV and $N_{\rm H}=5.9\times 10^{19}~\rm cm^{-2}$
(Paerels, Heise, \& van Teeseling 1994), and those for the {\it EUVE\/} SW spectrum are $kT
=18$~eV and $N_{\rm H}=8.8\times 10^{19}~\rm cm^{-2}$ (Mauche, Paerels, \& Raymond 1995).
The inability to derive consistent results from grating observations of the brightest polar
should warn us not to take too seriously the parameters---both direct ($kT$, $N_{\rm H}$,
solid angle) and inferred (fractional emitting area, bolometric luminosity)---derived from
such simple model fitting. A much better approach is to look for {\it trends\/} in the
spectral parameters of a {\it sample\/} of systems analyzed in a consistent manner, and to
investigate the sensitivity of the derived parameters to the model assumptions. Such is
just the purpose of this presentation.

\section{EUVE Spectra}

Until the launch of {\it AXAF\/} later this year, there is a single satellite capable of
dispersive spectroscopy of the soft spectral component of magnetic CVs: the {\it Extreme
Ultraviolet Explorer\/} ({\it EUVE\/}; Bowyer \& Malina 1991; Bowyer et al.\ 1994). The
salient features of {\it EUVE\/}'s SW spectrometer are its 70--180~\AA \ bandpass, its
0.5~\AA\ spectral resolution, and its relatively small effective area ($\approx 2~\rm
cm^2$ at 100~\AA ). The last attribute means that bright targets and long integrations
are required to obtain useful EUV spectra, and integrations of 50--100 kiloseconds are
consequently typical. Such long integrations assure that all binary orbital phases
are sampled, but the low count rates typically do not allow studies of the orbital phase
dependence of the spectra. While the width of the SW bandpass is nominally a factor of
2.6, it is typically effectively much narrower because of photoelectric absorption of EUV
photons by material within the binary (e.g., the accretion stream and column) and the
interstellar medium between the source and Earth; unit optical depth is reached for a
column density of $10^{18}$, $10^{18.5}$, $10^{19}$, $10^{19.5}$, and $10^{20}~\rm cm^{-2}$
at $\sim 400$, 250, 150, 100, and 65~\AA , respectively. 

At the present time (1998 August), there are 17 magnetic CVs with {\it EUVE\/} spectra
in the public archive (for a general discussion of these and other {\it EUVE\/} spectra,
see Craig et al.\ 1997). Only 2 of these 17 systems are intermediate polars (EX~Hya and 
PQ~Gem), and since papers have been published on both of these systems (Hurwitz et al.\
1997 and Howell et al.\ 1997, respectively), their spectra will not be discussed here. Of
the 15 polars, only 11 have ``useful'' spectra, and details of the relevant observations
of these 11 systems are collected in Table~1. The columns in that table are as follows.
The second column is the UT date of the start of the observation. The third column is the
Primbsch/deadtime corrected exposure time for the SW image. The fourth column indicates
whether the spectrum was dithered (delightfully, to ``dither,'' is to ``shiver'' or
``tremble'') on the face of the detector to eliminate the detector fixed-pattern noise;
well-exposed non-dithered spectra have non-statistical errors in the derived flux densities
which artificially increase the $\chi ^2$ of fits to the data. The fifth column reports the
maximum signal-to-noise ratio of the data in 0.54~\AA \ bins. Finally, for completeness,
the sixth column supplies a reference to a previous work with some discussion of the {\it
EUVE\/} spectrum of each source. For all of the sources except AM~Her, the given reference
deals with the same spectrum as that discussed herein; the reference for AM~Her is for the
paper on the original (1993 September) undithered observation of that source. Like AM~Her,
QS~Tel has been observed repeatedly by {\it EUVE\/}, and for both of these sources we have
extracted from the archive the longest single exposure.

% ------------------------------Table 1-----------------------------------------
\vspace{-4mm}
\begin{table*}
\begin{center}
\begin{tabular}{lccccc}
\multicolumn{6}{c}{TABLE 1} \\
\multicolumn{6}{c}{\vspace{-3mm}} \\
\multicolumn{6}{c}{Journal of Observations} \\
\multicolumn{6}{c}{\vspace{-3mm}} \\
\tableline \tableline
\multicolumn{6}{c}{\vspace{-3mm}} \\
& Start Date& Exposure& & & \\
Source& (UT m/d/y h:m)& (ksec)& Dithered?& S/N& Ref.$^a$\\
\multicolumn{6}{c}{\vspace{-3mm}} \\
\tableline
\multicolumn{6}{c}{\vspace{-3mm}} \\
AM Her\dotfill &   03/08/95~~12:19&   123.3&  Yes& 46!!& 1\\
AN UMa\dotfill &   02/27/93~~22:15&   !41.1&   No& !2.5& 2\\
AR UMa\dotfill &   12/14/96~~09:25&   !93.7&  Yes& 22!!&  \\
BL Hyi\dotfill &   10/30/95~~07:37&   !39.8&   No& !8!!& 3\\
EF Eri\dotfill &   09/05/93~~13:42&   !95.7&   No& !7!!& 4\\
QS Tel\dotfill &   10/06/93~~07:51&   !69.5&   No& 13!!& 5\\
RE J1149$+$284 &   12/26/94~~06:06&   114.3&  Yes& !2.5&  \\
RE J1844$-$741 &   08/17/94~~13:58&   134.6&   No& !8!!&  \\
UZ For\dotfill &   01/15/95~~20:38&   !78.5&  Yes& !7!!& 2\\
VV Pup\dotfill &   02/07/93~~21:24&   !43.6&   No& !5!!& 6\\
V834 Cen\dotfill & 05/28/93~~03:06&   !41.3&   No& !8!!& 2\\
\multicolumn{6}{c}{\vspace{-3mm}} \\
\tableline
\end{tabular}
\end{center}
\vspace{0 mm}
\mbox{\quad $^a$References: 1: Paerels et al.\ 1996b; 2: Warren 1998; 3: Szkody et al.\  }
\mbox{\quad \phantom{$^a$}1997; 
                4: Paerels et al.\ 1996a; 5: Rosen et al.\  1996; 6: Vennes et al.\ 1995.}
\end{table*}
% ------------------------------------------------------------------------------

For the record, the reduction of the archival data was accomplished as follows. The SW
image was extracted from the FITS data file, while the effective exposure time and
wavelength parameters were extracted from the FITS header. The centerline of the spectrum
was determined by forming a projection of the SW image onto the imaging axis. The source
region was taken to be this centerline $\pm 10$ lines (e.g., lines 137--157), while the
background region was taken to be 84 lines above and below the source region beyond a gap
of 10 lines (e.g., lines 44--127 and 167--250). The source and background spectra are the
sum of the counts in these regions within each wavelength bin, and the net spectra and
errors were calculated accordingly after binning in wavelength by a factor of 8 (from
$\Delta\lambda = 0.0674$~\AA \ to 0.539~\AA ). This wavelength binning matches the spectral
resolution of the SW detector, hence any intrinsically narrow absorption or emission
features will appear predominantly in one wavelength bin. It is at this binning that the
signal-to-noise ratio values shown in Table~1 were derived. The two systems in that table
with {\it peak\/} signal-to-noise ratios below 3 (AN~UMa and RE 1149) were not considered
further.

% ------------------------------Figure 1a---------------------------------------
\begin{figure}
% bounding box in ps file modified to account for label (-36 in bbllx and -23 in bblly):
% BoundingBox: 189 243 503 719 -> 153 220 503 719
\psfig{file=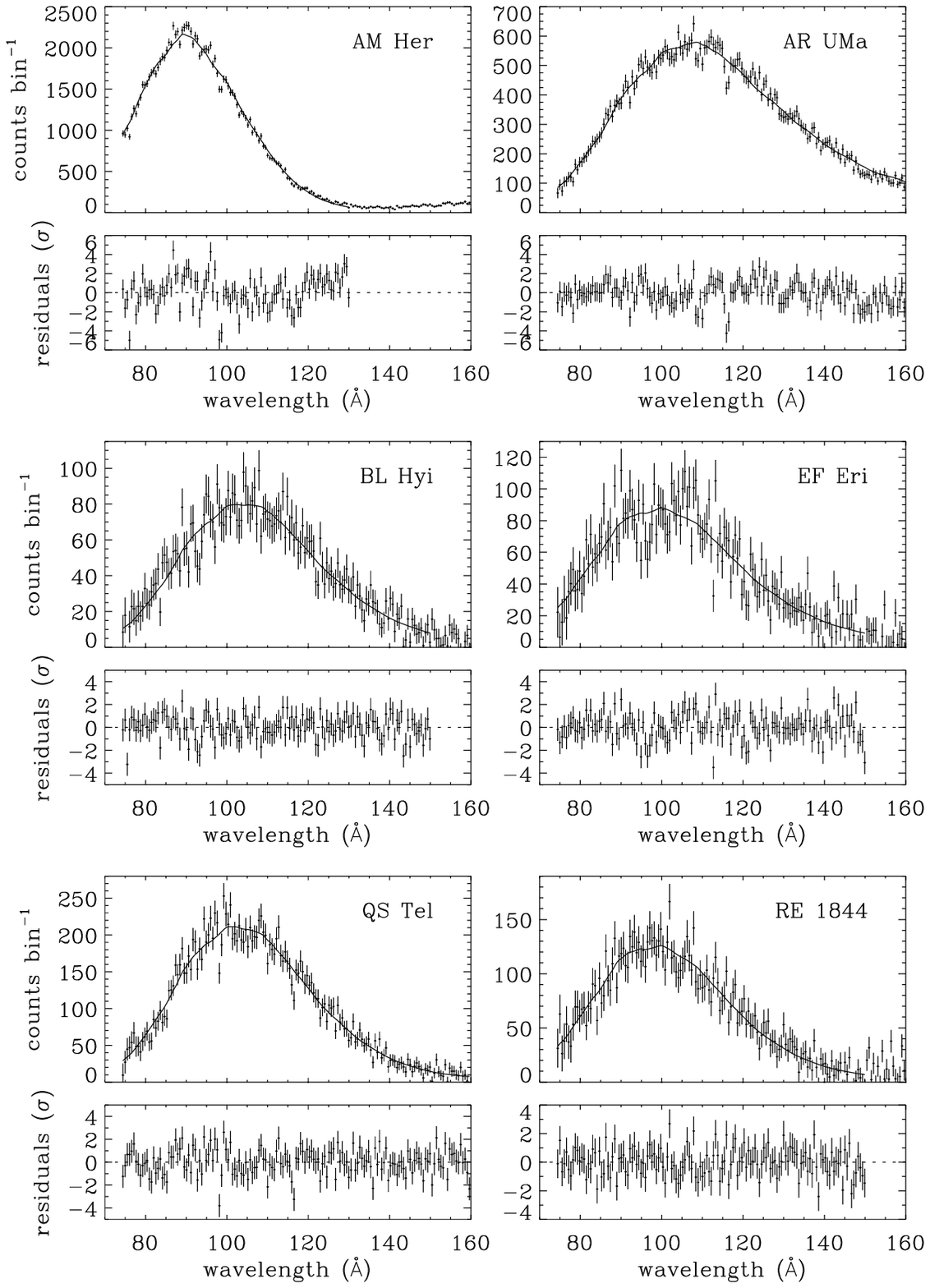}
\end{figure}
% ------------------------------------------------------------------------------

The resulting spectra (in counts per 0.54~\AA \ bins versus wavelength) of the 9 surviving
sources are shown in Figure~1. The shape of these spectra naturally mimic the shape of the
effective area curve, which peaks at 100~\AA \ and falls off at both ends of the bandpass.
Of the count distributions shown in Figure~1, that of AR~UMa is the softest, as it peaks
at $\sim 110$~\AA \ and not only extends all the way down to 180~\AA\ in the SW channel,
but even manifests itself on the ``left'' end of the MW channel (150--350~\AA ). In
contrast, the count distribution of AM~Her is among the hardest of the sources shown, as
it peaks at $\sim 90$~\AA \ and falls off rapidly at longer wavelengths. Note, however,
that while the long-wavelength ($\lambda > 130$~\AA ) flux of AM~Her is small, it is not
zero; indeed, the count distribution {\it rises\/} longward of $\sim 140$~\AA . Given the
exponential form of photoelectric absorption, this apparent long-wavelength flux is almost
certainly due to ``contamination'' of the first-order spectrum by higher orders. While the
second- and third-order diffraction efficiencies of the {\it EUVE\/} spectrometers were
measured in the laboratory prior to launch, it is less clear that they were calibrated in
orbit. To minimize the possible uncertainties of the higher-order diffraction efficiencies,
we ignore the data longward of some wavelength where higher-order flux may dominate the
first-order flux. For AM~Her, we take this cutoff to be at 130~\AA ; for the other sources,
it ranges from 135 to 180~\AA . The short-wavelength limit of the spectra is fixed at
74~\AA ; shortward of that wavelength there is a rapid increase in the background.

% ------------------------------Figure 1b---------------------------------------
\begin{figure}
% bounding box in ps file modified to account for label (-36 in bbllx and -23 in bblly):
% BoundingBox:  189 243 503 719 ->  189 414 503 719 ->  153 391 503 719
\psfig{file=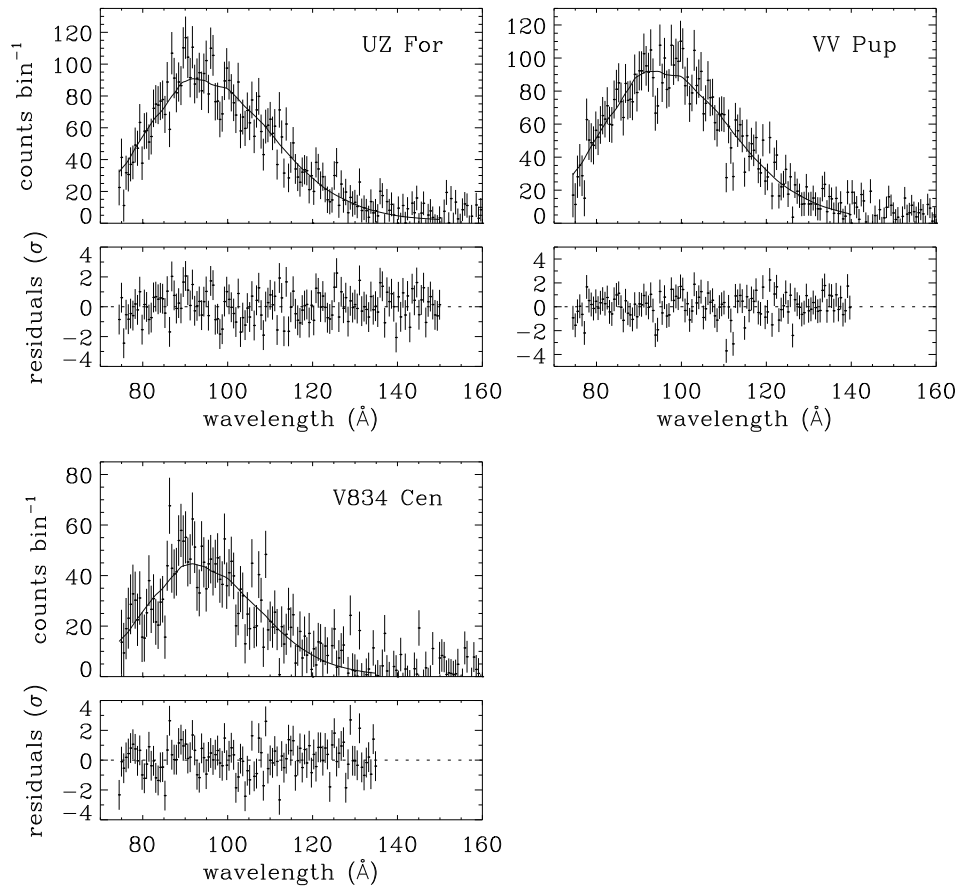}
\caption{{\it EUVE\/} spectra in counts per 0.54~\AA \ bin and the residuals relative to
the best-fit blackbody model.}
\end{figure}
% ------------------------------------------------------------------------------

To allow a quantitative assessment of the spectra shown in Figure~1, these data were fit
with three different spectral models---a blackbody, a pure-H stellar atmosphere, and a
solar abundance stellar atmosphere---all extinguished at long wavelengths by
photoelectric absorption. For the latter, we take the EUV absorption cross sections of
Rumph, Bowyer, \& Vennes (1994) for H\,{\sc i}, He\,{\sc i}, and He\,{\sc ii} with
abundances ratios of 1:0.1:0.01, as is typical for the diffuse interstellar medium. While
this choice for the abundance ratios is standard, it is decidedly non-trivial, since the
slopes of the absorption cross sections of the various ions differ somewhat in the EUV.
For the chosen ratios, the photoelectric opacity in the SW bandpass is dominated by
He\,{\sc i}, while for much more highly ionized gas (e.g., the accretion stream and
column), He\,{\sc ii} may dominate. Furthermore, partial covering may allow an excess of
EUV photons to escape the binary, but to be detected at Earth, these rogue photons must
still make their way through the ISM without getting clobbered.

\section{Blackbody Fits and General Comments}

The blackbody fits to the {\it EUVE\/} spectra are the simplest to calculate as well as to
describe, hence we begin with those. The fits of this model to the {\it EUVE\/} data and
the resulting residuals are shown in Figure~1, while the 68, 90, and 99\% confidence
contours are shown in Figure~2, and the 90\% confidence fit parameters ($kT$, $N_{\rm H}$,
solid angle, 70--140~\AA \ flux, bolometric flux, $\chi ^2$/dof) are listed in Table~2.
First consider the best-fit models and residuals shown in Figure~1.

Blackbodies may or may not be an accurate description of the intrinsic EUV spectra of
polars, but this model is smooth and hence its residuals reveal the presence of spectral
features such as lines and edges. The number of possible discrete transitions in the SW
bandpass is huge, but among the abundant elements, possible absorption edges include
N\,{\sc v}, O\,{\sc v}--{\sc vi}, Ne\,{\sc iv}--{\sc vi}, Mg\,{\sc iii}--{\sc v},
S\,{\sc vi}, Ar\,{\sc vi}--{\sc viii}, Ca\,{\sc v}--{\sc viii}, and Fe\,{\sc vi}--{\sc
viii}. Because of the high density of the white dwarf photosphere, there are in addition
to the ground-state edges (e.g., the O\,{\sc vi} $1s^22s$ edge at 89.8~\AA \ and the
Ne\,{\sc vi} $2s^22p$ edge at 78.5~\AA), edges from excited states of these ions (e.g.,
the O\,{\sc vi} $1s^22p$ edge at 98.3~\AA \ and the Ne\,{\sc vi} $2s2p^2$ edge at
85.2~\AA). O\,{\sc vi} edges were identified by Vennes et al.\ (1995) in the spectrum of
VV~Pup, Ne\,{\sc vi} edges were identified by Paerels et al.\ (1996b) in the 1993 September
spectrum of AM~Her, and the Ne\,{\sc vi} $2s2p^2$ edge was identified by Rosen et al.\
(1996) in the spectrum of QS~Tel. The edge in the QS~Tel spectrum is just visible in
Figure~1 as a discontinuous jump at 85~\AA \ in the residuals for this source, but the
putative edges of AM~Her and VV~Pup are less obvious. Perhaps the most obvious jump in the
residuals is manifest by V834~Cen at 85~\AA , again implicating Ne\,{\sc vi}. The problem
with detecting edges this way is that there are medium- and low-frequency residuals present
at some level in almost all of the spectra, even though the reduced $\chi ^2$ of the fits
listed in Table~2 indicate that for most of the sources the fits are acceptable.

It is more straightforward to detect discrete features in these spectra, since the spectral
binning is set to match the resolution of the SW instrument and because such features
are apparent almost regardless of the adopted spectral model. With the exception of AM~Her,
there are no sources with discrete residuals greater than $+3\sigma $ (i.e., emission
lines), while nearly all of the sources show discrete residuals less than $-3\sigma $
(i.e., absorption lines). The features with the highest significance in one spectral bin
are found in the residuals of AM~Her (76.1, 98.2~\AA ), AR~UMa (116.5~\AA ), and QS~Tel
(98.2, 116.5~\AA ). These are the very sources with the highest signal-to-noise ratio
spectra ($\rm S/N > 10$), suggesting the possibility that similar features could be
detected in all of the sources if the integrations were long enough. The 98~\AA \ feature
is identified as Ne\,{\sc viii} $2p$-$3d$ and was observed first by Paerels et al.\ (1996b)
in the 1993 September {\it EUVE\/} spectrum of AM~Her, and subsequently by Rosen et al.\
(1996) in the spectrum of QS~Tel; in the new dithered spectrum of AM~Her this feature is
so strong (and the signal-to-noise ratio so high) that it is readily apparent in the raw
data. The 116.5~\AA\ feature was observed first by Rosen et al.\ in QS~Tel and is
identified as Ne\,{\sc vii} $2s2p$-$2s3d$; we now identify this feature in AR~UMa as well.
Other reasonably narrow and apparently real absorption features are found in the residuals
of AR~UMa (108.7~\AA ), BL~Hyi (92.9~\AA ), EF~Eri (96.5~\AA ), VV~Pup (94.1~\AA), but
their identifications are uncertain.

% ------------------------------Figure 2----------------------------------------
\begin{figure}
% bounding box in ps file modified to account for label (-36 in bbllx and -23 in bblly):
% BoundingBox: 189 531 503 719 -> 153 508 503 719
\psfig{file=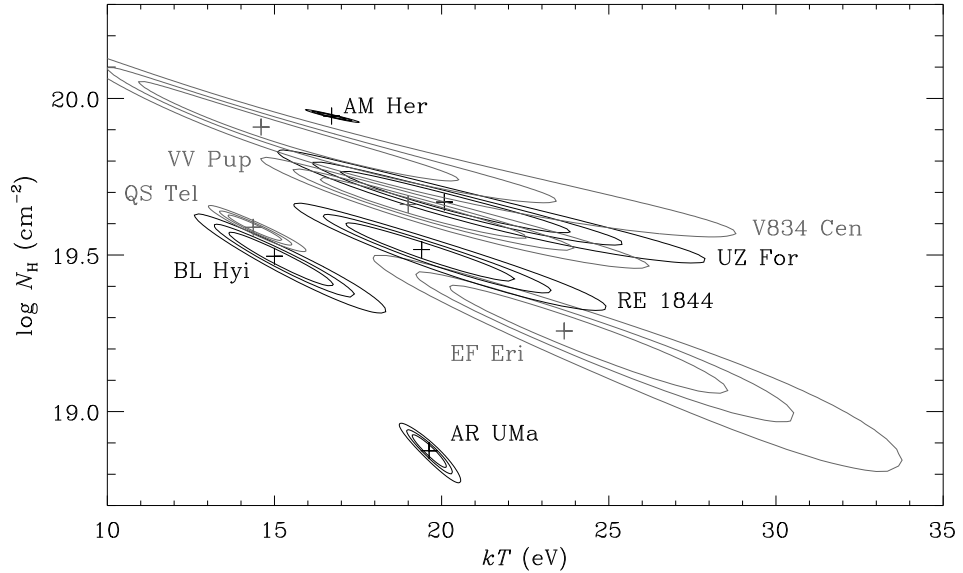}
\caption{68, 90, and 99\% confidence contours in the $kT$--$N_{\rm H}$ plane for the
blackbody model fits to the {\it EUVE\/} spectra.}
\end{figure}
% ------------------------------------------------------------------------------

The $\chi ^2$ surface for the blackbody fits to the {\it EUVE\/} data is shown in Figure~2,
which shows that within the 90\% confidence contours, the blackbody temperature ranges
between 13.4 and 20.3~eV (156--236 kK). On the orthogonal axis, the hydrogen column density
ranges from a low of $8\times 10^{18}~\rm cm^{-2}$ for AR~UMa to a high of $9\times
10^{19}~\rm cm^{-2}$ for AM~Her. If this value for $N_{\rm H}$ for AM~Her is physical and
not simply a parameterization of the data, most of the absorbing column must be ionized
and hence within the binary, since the neutral hydrogen column density to this source is
$\approx 3\times 10^{19}~\rm cm^{-2}$ (G\"ansicke et al.\ 1998). Table~2 lists the
corresponding 90\% confidence parameters for these blackbody fits. Note that the reduced
$\chi^2 $ of the fits to AM~Her, AR~UMa, EF~Eri, and QS~Tel are not acceptable, so the fit
parameters should be taken only as indicative. For a given distance $d$ to a given source,
the tabulated values of the solid angle $\Omega = (r/d)^2$ and bolometric flux $L/4\pi
d^2$ can be used to derive such useful quantities as the fractional emitting area and the
bolometric luminosity.

% ------------------------------Figure 3----------------------------------------
\begin{figure}
% bounding box in ps file modified to account for label (-36 in bbllx and -23 in bblly):
% BoundingBox: 189 531 415 719 -> 153 508 415 719 -> 132 508 415 719
\psfig{file=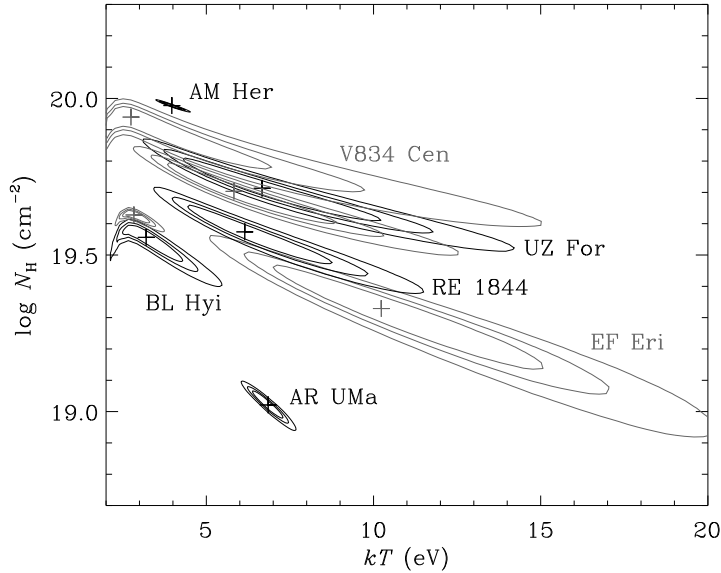}
\caption{68, 90, and 99\% confidence contours in the $kT$--$N_{\rm H}$ plane for the
pure-H stellar atmosphere model fits to the {\it EUVE\/} spectra.}
\end{figure}
% ------------------------------------------------------------------------------

\section{Pure-H Stellar Atmosphere Fits}

The second model used to fit the {\it EUVE\/} spectra was that of a pure-H, line-blanketed,
NLTE, $\log g = 8$ stellar atmosphere calculated with TLUSTY (Hu\-beny 1988). The $\chi
^2$ surface of the fits of these models to the data is shown in Figure~3. Notice that the
relative ordering of these Daliesque contours is very similar to that of the blackbody
fits, but that the temperatures are systematically much lower: within the 90\% confidence
contours, the effective temperature ranges between 2.4 and 7.5~eV (28--87 kK). The
hydrogen column densities are higher than before, but typically by only 10--20\%. The 90\%
confidence parameters for these fits are again listed in Table~2. Note that the reduced
$\chi ^2$ of these fits are essentially identical to those of the blackbody model, hence
at that level the pure-H stellar atmosphere model is just as acceptable a description of
the data.  However, because the EUV bump in the stellar atmosphere models contains a
relatively small fraction of the total luminosity, the biggest change between these models
is the solid angle, which is now larger by a factor of $\sim 10^2$--$10^4$. Indeed, in
some cases (e.g., AM~Her, QS~Tel) the derived solid angle is so large that it completely
excludes the stellar atmosphere model: since $R_{\rm wd}<10^9$ cm and $d>75$ pc, the solid
angle must be less than $\Omega _{\rm wd} = (R_{\rm wd}/d)^2 = 2\times 10^{-23}$. In other
cases, the implied UV flux density will likely exceed the measured value, but such a
constraint typically requires that we have simultaneous EUV and UV measurements, which is
seldom the case.

% ------------------------------Figure 4----------------------------------------
\begin{figure}
% bounding box in ps file modified to account for label (-36 in bbllx and -23 in bblly):
% BoundingBox: 258 531 434 719 -> 222 508 434 719 -> 201 508 434 719
\psfig{file=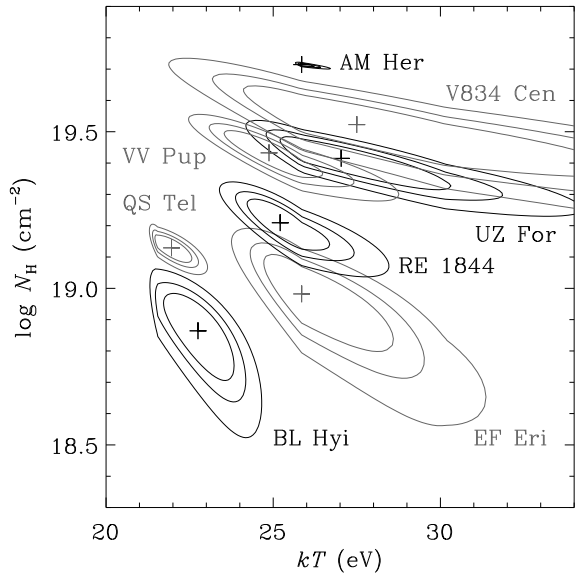}
\caption{68, 90, and 99\% confidence contours in the $kT$--$N_{\rm H}$ plane for the
solar abundance stellar atmosphere model fits to the {\it EUVE\/} spectra. The kinks
in the contours are the result of interpolating in a rather sparse grid of models.}
\end{figure}
% ------------------------------------------------------------------------------

\section{Solar Abundance Stellar Atmosphere Fits}

The third and final model used to fit the {\it EUVE\/} spectra was that of a solar
abundance stellar atmosphere; specifically, the ``un-illuminated'' solar abundance model
atmospheres of van Teeseling, Heise, \& Paerels (1994). The $\chi^2$ surface of the fits
of these models to the data is shown in Figure~4. Table~2 again lists the nominal 90\%
confidence parameters values for these fits, but note that because the reduced $\chi ^2$
of the fits are unacceptably large the parameters should be understood only to be
indicative. With this limitation in mind, we see that the effective temperatures are
higher, and the hydrogen column density lower compared to the blackbody and pure-H
stellar atmosphere model fits; within the nominal 90\% confidence contours, the effective
temperature ranges between 22.8 and 27.4~eV (265--318 kK), while the hydrogen column
density ranges from a (rather unlikely) low of $<6\times 10^{17}~\rm cm^{-2}$ for AR~UMa
to a high of $5\times 10^{19}~\rm cm^{-2}$ for AM~Her. This range of parameters is pleasing
for two reasons. First, the inferred hydrogen column densities are lower than for the
previous models and are more likely to be consistent with the interstellar values. Second,
the inferred effective temperatures are now high enough that the accretion region may be
capable of producing the soft X-ray fluxes observed by {\it ROSAT\/} (e.g., Beuermann \&
Burwitz 1995).

% ------------------------------Table 2-----------------------------------------
\begin{figure}
\psfig{file=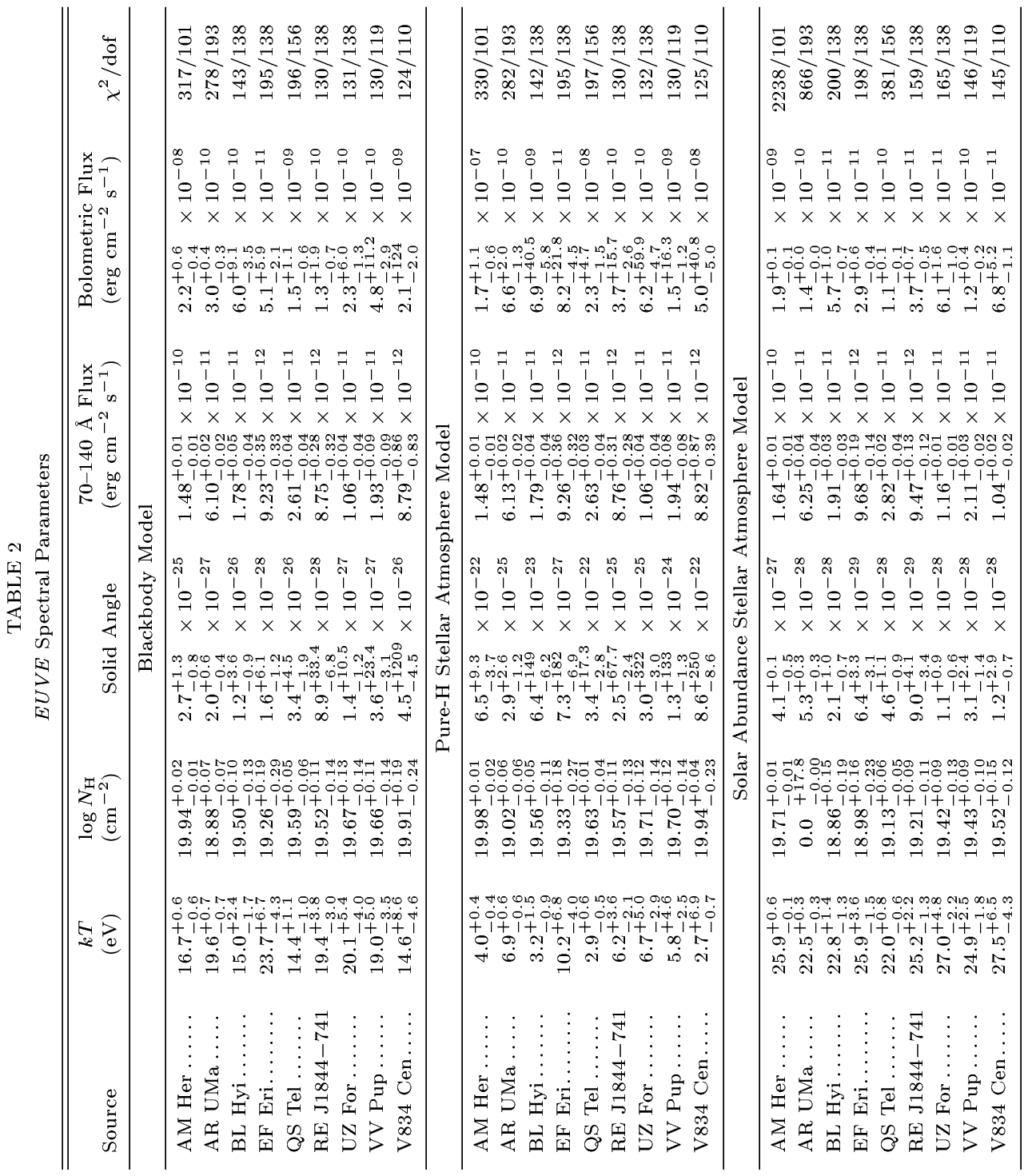,angle=90}
\end{figure}
% ------------------------------------------------------------------------------

Despite these attractive aspects of the solar abundance stellar atmosphere models, they
are in every case an unacceptable description of the data because of their strong O\,{\sc
vi} absorption edges. The simplest way to remedy this discrepancy is to reduce the O
abundance, but there is no other compelling evidence that the material accreted by the
white dwarf in polars is significantly underabundant in this element; AM~Her for one
certainly has a strong O\,{\sc vi} $\lambda 1035$ emission line in the FUV (Mauche \&
Raymond 1998). A much more natural and physical explanation of the weakness of the O\,{\sc
vi} absorption edges is the irradiated stellar atmosphere model (Williams, King, \& Booker
1987; van Teeseling, Heise, \& Paerels 1994). In that model, irradiation of the white
dwarf photosphere by hard X-rays results in a temperature inversion above the photosphere
and a flattening of the run of temperature with optical depth within the photosphere. If
the temperature profile in the photosphere is flat where the edges and lines form, their
strength will be significantly decreased. van Teeseling et al.\ found that an irradiated
stellar atmosphere with an effective temperature of $kT<9$~eV ($<100$~kK) fit the {\it
EXOSAT\/} TGS spectrum of AM~Her as well as or better than a blackbody with $kT=23$~eV
(270~kK). Note, however, that their model requires that more than 96\% of the soft
X-ray luminosity is due to reprocessing; this leaves little or no room for direct kinetic
heating of the photosphere, the favored solution of the famous soft X-ray problem.
Furthermore, even the irradiated stellar atmosphere models have strong edges shortward of
the {\it EUVE\/} bandpass, so it is not entirely clear that the observed soft X-ray fluxes
can be produced.

\section{Summary}

We have found that, of the blackbody, pure-H stellar atmosphere, and solar abundance
stellar atmosphere models, the blackbody model provides the best phenomenological
description of the {\it EUVE\/} 70--180~\AA \ spectra of polars. Inadequacies of this
model include the weak absorption edges of Ne\, {\sc vi} and the absorption lines of Ne\,
{\sc vii} and Ne\, {\sc viii} apparent in the residuals of the sources with the highest
signal-to-noise ratio spectra, and the likely inability of these moderately soft
blackbodies to produce the observed soft X-ray fluxes. The untested irradiated solar
abundance stellar atmosphere model is likely a better overall description of the EUV/soft
X-ray spectra of polars, but better models (which include, e.g., absorption lines as well
as edges) and better data (e.g., high signal-to-noise ratio phase-resolved {\it AXAF\/}
3--140~\AA \ LETG spectra) are required before significant progress can be made in our
understanding of the accretion region of magnetic CVs.

\acknowledgments
The author is pleased to acknowledge I.~Hubeny for his guidance in the proper use of the
TLUSTY and SYNSPEC suite of programs, A.~van Teeseling for generously providing his grid
of solar abundance stellar atmosphere models, B.~G\"ansicke for his assistance spot
checking his and TLUSTY's pure-H stellar atmosphere models, and the hospitality of the
students and staff---particularly A.~Evans, R.~Jeffries, T.~Naylor, and J.~Wood---at Keele
University where this work was completed. This work was performed under the auspices of
the U.S.~Department of Energy by Lawrence Livermore National Laboratory under contract
No.~W-7405-Eng-48.

% ------------------------------------------------------------------------------


\begin{references}
\reference Beuermann, K., \& Burwitz, V. 1995, in Cape Workshop on Magnetic Cataclysmic
           Variables, ed.\ D.~A.~H.\ Buckley \& B.~Warner (San Francisco: ASP), 99
\reference Bowyer, S., \& Malina, R.~F. 1991, in Extreme Ultraviolet Astronomy, ed.\
           R.~F.\ Malina \& S.~Bowyer (New York: Pergamon), 397
\reference Bowyer, S., et al. 1994, \apjs , 93, 569
\reference Craig, N., et al. 1997, \apjs, 113, 131
\reference G\"ansicke, B.~T., Hoard, D.~W., Beuermann, K., Sion, E.~M., \& Szkody, P.
           1998, \aap, in press
\reference Heise, J., et al. 1984, Phys.\ Scripta, T7, 115
\reference Howell, S.~B., et al. 1997, \apj, 485, 333
\reference Hubeny, I. 1988, Computer Phys.\ Comm., 52, 103
\reference Hurwitz, M., Sirk, M., Bowyer, S., \& Ko, Y.-K. 1997, \apj, 477, 390
\reference Mauche, C.~W., Paerels, F.~B.~S., \& Raymond, J.~C. 1995, in Cape Workshop
           on Magnetic Cataclysmic Variables, ed.\ D.~A.~H.\ Buckley \& B.~Warner
           (San Francisco: ASP), 298
\reference Mauche, C.~W., \& Raymond, J.~C. 1998, \apj, 505, in press
\reference Paerels, F., Heise, J., \& van Teeseling, A. 1994, \apj, 426, 313
\reference Paerels, F., Hur, M.~Y., \& Mauche, C.~W. 1996a, in Astrophysics in the
           Extreme Ultraviolet, ed.\ S.~Bowyer \& R.~F.\ Malina (Dordrecht: Kluwer), 309
\reference Paerels, F., Hur, M.~Y., Mauche, C.~W., \& Heise, J. 1996b, \apj,
           464, 884
\reference Rosen, S.~R., et al. 1996, \mnras, 280, 1121 (note that the units of the y
           axis of Fig.~6 of this paper should be
           $10^{-3}~\rm photons~cm^{-2}~s^{-1}~\AA ^{-1}$)
\reference Rumph, T., Bowyer, S., \& Vennes, S. 1994, \aj, 107, 2108
\reference Szkody, P., Vennes, S., Sion, E.~M., Long, K.~S., \& Howell, S.~B. 1997,
           \apj, 487, 916 (note that the units of the y axis in the upper panels of
           Fig.~2 of this paper should be $10^{-12}~\rm erg~cm^{-2}~s^{-1}~\AA ^{-1}$)
\reference van Teeseling, A., Heise, J., \& Paerels, F. 1994, \aap, 281, 119
\reference Vennes, S., Szkody, P., Sion, E.~M., \& Long, K.~S. 1995, \apj, 445,
           921
\reference Warren, J.~K. 1998, PhD thesis, Physics Department, UC Berkeley
\reference Williams, G.~A., King, A.~R., \& Booker, J.~R.~E. 1987, \mnras, 266, 725
\end{references}
\end{document}